\documentclass{ws-procs9x6}

\newcommand{\be}{\begin{equation}}
\newcommand{\ee}{\end{equation}}
\newcommand{\beq}{\begin{eqnarray}}
\newcommand{\eeq}{\end{eqnarray}}

\begin{document}

\title{N and N to $\Delta$ transition form factors from Lattice QCD}

\author{C. Alexandrou}

\address{Department of Physics, University of Cyprus,\\
P.O. Box 20537, CY-1678 Nicosia, Cyprus\\
E-mail: alexand@ucy.ac.cy}



\begin{abstract}
We present recent lattice QCD results on nucleon form factors  
and N to $\Delta$
transition form factors. We predict the parity violating asymmetry in 
N to $\Delta$ and check the off-diagonal Goldberger-Treiman relation.
\end{abstract}


\bodymatter

\vspace{0.5cm}


We present the evaluation, within lattice QCD, of fundamental physical 
quantities of the  nucleon-$\Delta$ system. On the theoretical side, 
providing a complete set of 
form factors and coupling constants constitutes an important input for
model builders and for fixing the parameters of chiral effective theories.
On the experimental side, there is ongoing effort to measure accurately
these quantities. Examples are the  recent polarization experiments,
 of the electric, $G_E$, and magnetic, $G_M$, nucleon form factors, 
and the accurate measurements on the electric and scalar quadrupole multipoles
  and the magnetic dipole in N to $\Delta$ transition as well as
the ongoing experiment to measure the parity violating asymmetry 
in N to $\Delta$. 
Using state-of-the-art-lattice techniques we obtain results with small 
statistical errors for pion masses, $m_\pi$, in the range 600-360 MeV. We use 
two flavors of dynamical Wilson fermions and domain wall valence quarks (DWF)
on MILC configurations to study the role of  pion cloud contributions.
Results obtained with dynamical Wilson fermions and  DWF are in agreement 
showing that lattice artifacts are under control.

\begin{figure}

\begin{minipage}{5.5cm}
{\mbox{\includegraphics[height=6.2cm,width=5.4cm]{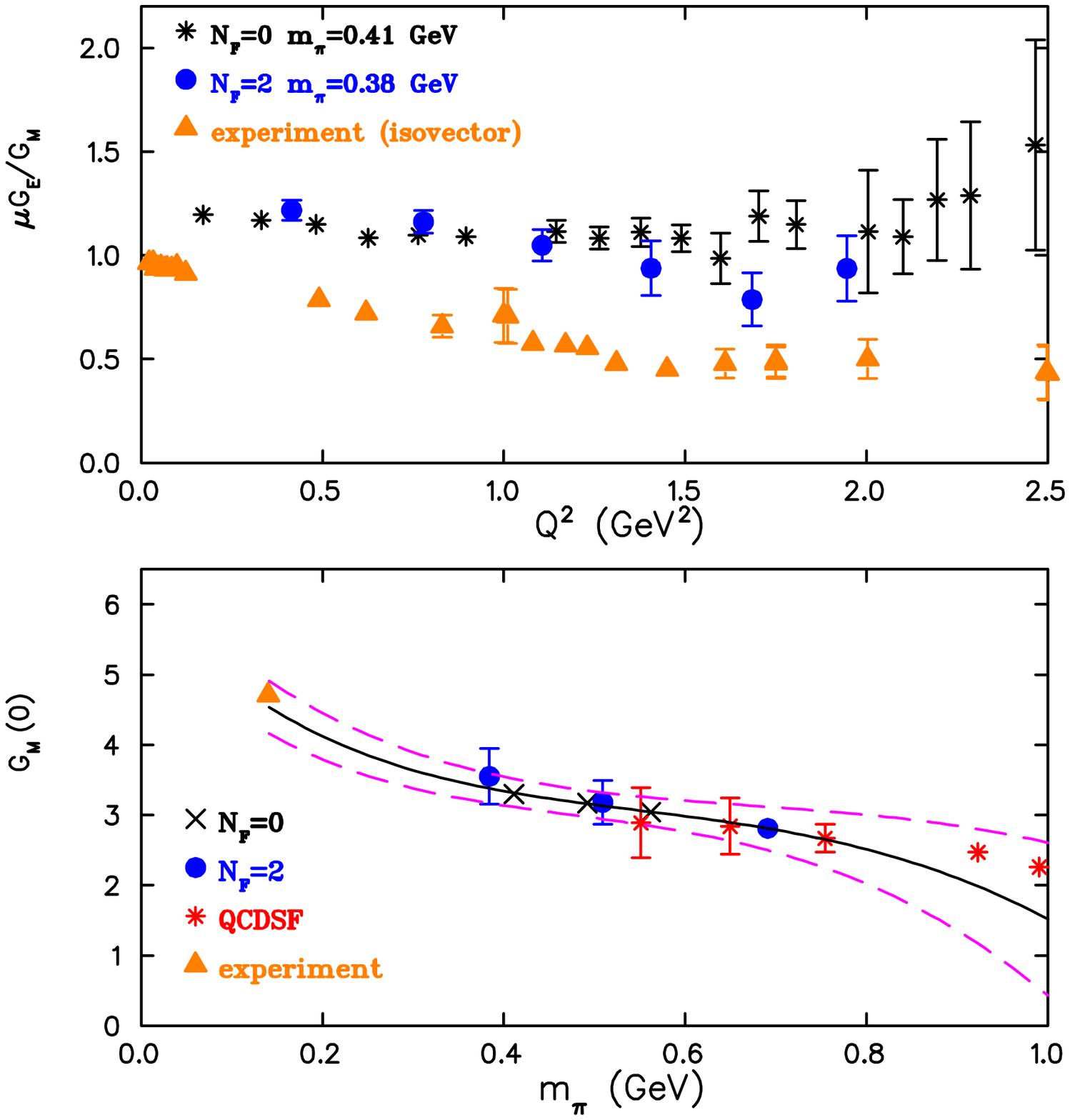}}} 
\vspace*{-0.2cm}
\caption{Top: $\mu G_E/G_M$   with $\mu=4.71$ versus $Q^2=-q^2$ for 
the lightest pion for quenched and dynamical Wilson fermions.
Bottom: The magnetic moment versus $m_\pi$.}
\label{fig:nucleonff}
\eject
\end{minipage}
\begin{minipage}{5.8cm}
\vspace*{-0.5cm}
{\mbox{\includegraphics[height=6.2cm,width=5.4cm]{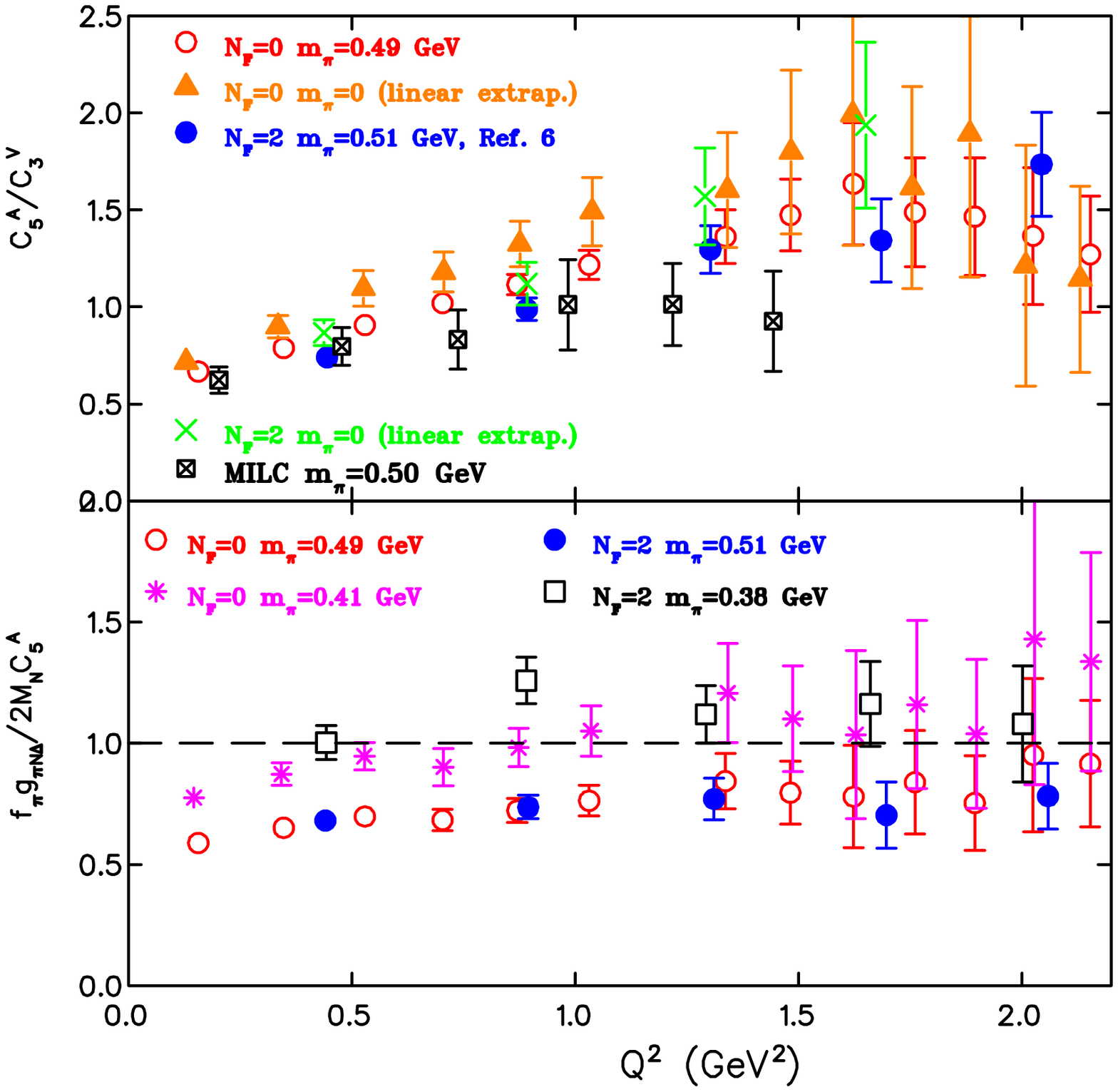}}} 
\caption{Top: $C_5^A/C_3^V$  versus $Q^2$ for Wilson and 
DWF fermions at $m_\pi\sim 500$ MeV and, 
 at the physical limit for Wilson fermions.
Bottom: The ratio
$R_{GT}$ for Wilson fermions versus $Q^2$ for the two smallest pion masses.}
\label{fig:asymmetry}
\end{minipage}
\vspace*{-0.3cm}
\end{figure}

In this work only the isovector  nucleon form factors are evaluated since
 isoscalar contributions involved quark loops
that are technically difficult to calculate. 
In Fig.~\ref{fig:nucleonff} we display the momentum dependence of the 
 ratio of the isovector electric to magnetic form factors
for the lightest pion mass namely
410 MeV in the quenched theory and 380 MeV for 
dynamical Wilson fermions~\cite{NN}.
The lattice results are in agreement but higher than experiment. 
Given the weak quark mass dependence of this ratio  
for the quark masses used in this work  a linear extrapolation
in $m_\pi^2$ to the physical limit fails to reproduce experiment. 
On the other hand, the isovector magnetic
moments extracted from lattice results using a dipole Ansatz are well
described by
chiral effective theory~\cite{chiral} that includes explicitly the $\Delta$.
As can be seen in Fig.~\ref{fig:nucleonff} the extrapolated value of
the magnetic moment is
in agreement with experiment. This is consistent with the fact that lattice
results are closer to experiment for the magnetic than for the electric 
form factor.  

We evaluate the three electromagnetic Sachs 
and four axial Adler form factors  for the N to $\Delta$
transition.
We show in 
Fig.~\ref{fig:asymmetry} the 
the ratio $C_5^A/C_3^V$, which is the analogue of   $g_A/g_V$ and
determines, to a first approximation, the parity violating 
asymmetry~\cite{axial} .
The off-diagonal 
Goldberger-Treiman relation
implies that the
 ratio $R_{GT}=\frac{f_\pi g_{\pi N\Delta}(Q^2)}{2M_N C_5^A(Q^2}$
is one, where
$g_{\pi N\Delta}(Q^2)$ is determined
from the matrix element of the pseudoscalar density 
$<\Delta^+|\bar{\psi}(x)\gamma_5\frac{\tau^3}{2}\psi(x)|p>$
and $f_\pi$ is the pion decay constant. As can be seen in 
Fig.~\ref{fig:asymmetry} this ratio approaches 
unity  as $m_\pi$ decreases 
from $\sim 500$ MeV to $\sim 410 (380)$ MeV  for quenched (dynamical) Wilson
fermions.

\bibliographystyle{ws-procs9x6}
\bibliography{ws-pro-sample}

\end{document}